\documentclass[]{aa}  

\usepackage{graphicx}
\usepackage{txfonts}
\usepackage{graphicx}
\usepackage{natbib}
\newcommand{\ltsima} {$\; \buildrel < \over \sim \;$}
\newcommand{\gtsima} {$\; \buildrel > \over \sim \;$}
\newcommand{\lta} {\lower.5ex\hbox{\ltsima}}
\newcommand{\gta} {\lower.5ex\hbox{\gtsima}}

\newcommand{\kms}{km\ s$^{-1}$}

\begin{document} 

\title{Outflows of hot molecular gas in ultra-luminous infra-red galaxies mapped with VLT-SINFONI}
\titlerunning{Outflows of hot molecular gas in ULIRGs}

   \author{B.\,H.\,C. Emonts
          \inst{1,2}
          \and L. Colina
          \inst{1}
          \and J. Piqueras-L\'{o}pez
          \inst{1}
          \and S. Garcia-Burillo
          \inst{3}
          \and M. Pereira-Santaella
          \inst{4}
          \and S. Arribas
          \inst{1}
          \and A.Labiano
          \inst{1}
          \and A. Alonso-Herrero
          \inst{1}
          }

   \institute{Centro de Astrobiolog\'{i}a (INTA-CSIC), Ctra de Torrej\'{o}n a Ajalvir, km 4, 28850 Torrej\'{o}n de Ardoz, Madrid, Spain\\
              \email{bjornemonts@gmail.com}
         \and
National Radio Astronomy Observatory, 520 Edgemont Road, Charlottesville, VA 22903
         \and
Observatorio Astron\'{o}mico Nacional (OAN), Observatorio de Madrid, Alfonso XII, 3, 28014, Madrid, Spain
         \and
Department of Physics, University of Oxford, Keble Road, Oxford, OX1 3RH, UK
             }

   \date{}

 
  \abstract{We present the detection and morphological characterization of hot molecular gas outflows in nearby ultra-luminous infra-red galaxies (ULIRGs), using the Spectrograph for Integral Field Observations in the Near Infrared (SINFONI) on the Very Large Telescope (VLT). We detect outflows observed in the 2.12$\mu$m H$_{2}$ 1-0\,S(1) line for three out of four ULIRGs that we analyzed, namely IRAS\,12112+0305, IRAS\,14348-1447, and IRAS\,22491-1808. The outflows are mapped on scales of $0.7-1.6$ kpc, show typical outflow velocities of 300$-$500 \kms, and appear to originate from the nuclear region. The outflows comprise hot molecular gas masses of M$_{\rm H_2\,(hot)}$ $\sim$ 6$-$8\,$\times$10$^{3}$ M$_{\odot}$. Assuming a hot-to-cold molecular gas mass ratio of 6\,$\times$\,10$^{-5}$, as found in nearby luminous infrared galaxies, the total (hot+cold) molecular gas mass in these outflows is expected to be M$_{\rm H_2\,(tot)}$ $\sim$ 1 $\times$ 10$^{8}$ M$_{\odot}$. This translates into molecular mass outflow rates of \.{M}$_{\rm H_2\,(tot)} \sim 30-85$ M$_{\odot}$ yr$^{-1}$, which is a factor of a few lower than the star formation rate in these ULIRGs. In addition, most of the outflowing molecular gas does not reach the escape velocity of these merger systems, which implies that the bulk of the outflowing molecular gas is re-distributed within the system and thus remains available for future star formation. The fastest H$_{2}$ outflow is seen in the Compton-thick AGN of IRAS\,14348-1447, reaching a maximum outflow velocity of $\sim$900 \kms. Another ULIRG, IRAS\,17208-0014, shows asymmetric H$_{2}$ line profiles different from the outflows seen in the other three ULIRGs. We discuss several alternative explanations for the line asymmetries in this system, including a very gentle galactic wind, internal gas dynamics, low-velocity gas outside the disk, or two superposed gas disks. We do not detect the hot molecular counterpart to the outflow previously detected in CO(2-1) in IRAS\,17208-0014, but we note that our SINFONI data are not sensitive enough to detect this outflow if it has a small hot-to-cold molecular gas mass ratio of $\lesssim$9\,$\times$\,10$^{-6}$.}

   \keywords{Galaxies: interactions — Galaxies: individual: IRAS 12112+0305 — Galaxies: individual: IRAS 14348-1447 — Galaxies: individual: IRAS 22491-1808 — Galaxies: individual: IRAS 17208-0014 — ISM: jets and outflows}

   \maketitle
%

\section{Introduction}

Ultra-luminous infra-red galaxies (ULIRGs) are dust-enshrouded galaxies with massive starbursts and often a deeply buried active galactic nucleus (AGN), which may have been triggered by a gas-rich galaxy merger \citep{san96}. ULIRGs emit the bulk of their radiation at mid- and far-IR wavelengths ($L_{\rm IR} \ge 10^{12}L_{\odot}$), where light from the starburst and AGN is absorbed and re-radiated by dust. While ULIRGs are rare in the nearby Universe and contribute at most a few percent of the infrared emission at low redshifts \citep{soi91}, they are major contributors to the star-formation activity of the Universe at high redshifts \citep[e.g.,][]{flo05,per05,cap06,mag13,nov17}. Thus, studying the evolution of ULIRGs at low redshifts, where we can resolve the physical processes that drive these systems, is crucial for understanding the dusty, IR-luminous phase in the evolution of galaxies throughout the Universe.

A crucial aspect in the evolution of ULIRGs has been the discovery of massive gas outflows driven by starburst or AGN activity. These outflows have been extensively studied in both ionized and neutral gas \citep[e.g.,][]{hec90,leh96,rup11,wes12,bel13,rup13a,rup15,mor13sci,rod13,fer15,arr14,caz14,caz16}. With the discovery of outflows of neutral gas also came the realization that the mass, momentum, and energy of galactic winds and AGN-driven outflows are much higher in the neutral than in the ionized phase \citep{mor05,emo05,rup13a,mah13}. However, the full impact of outflows in ULIRGs has only been revealed with the discovery of massive outflows of molecular gas. 

Molecular hydrogen is the raw fuel for star formation, and studies have revealed that molecular outflows can halt star formation, at least in the central regions of ULIRGs. These studies focused on the coldest phase of molecular outflows, by using strong tracers like CO and OH 119$\mu$m \citep[e.g.,][]{fer10,fer15,fis10,chu11,stu11,spo13,vei13,das12,das14,cic14,sak14,gar15,aal15,per16,gon17}. Direct measurements of molecular hydrogen in outflows rely on detecting warm and hot H$_{2}$ emission in the mid- to near-IR \citep[e.g.,][]{val99,ogl10,das11,gui12}. Because these H$_{2}$ lines are relatively weak compared to CO, studies of H$_{2}$ outflows are limited to low redshift galaxies \citep[e.g.,][]{wer93,tad14,hil14}, and only few studies exist that provide information on the geometry of spatially extended H$_{2}$ outflows \citep{rup13b,emo14,das15,per16}. However, the detection of H$_{2}$ in massive galaxies at $z \sim 2$ with the Spitzer Space Telescope shows the potential for studying the hot molecular Universe with the James Webb Space Telescope (JWST) \citep[][]{ogl12,gui15}. 

In this paper, we study the complex kinematics of the hot molecular H$_{2}$ gas in a sample of four nearby ULIRGs ($0.04<z<0.08$). By performing a multi-Gaussian decomposition of the 2.1218\,$\mu$m H$_{2}$\,1-0S(1) line, we discovered that three of these systems show evidence for an outflow of hot molecular gas, which we resolve both spatially and kinematically with sub-kpc scale resolution.

\section{Sample and data}

We used near-IR data from four ULIRGs at 0.043\,$\le$\,$z$\,$\le$0.083 obtained with the Spectrograph for Integral Field Observations in the Near Infrared (SINFONI) on the Very Large Telescope (VLT). These are IRAS 12112+0305, IRAS 14348-1447, IRAS 17208-0014, and IRAS 22491-1808. They are the four closest ULIRGs from a representative sample that was previously published by \citealt{piq12} (hereafter PL12). Three of the four ULIRGs consist of two galaxies that are in a pre-coalescent stage of the merger, with the individual nuclei still clearly separated by at least several kpc. Therefore, our sample of four ULIRGs consists of seven (merging) galaxies. 

\begin{figure}
\centering
\includegraphics[width=0.90\hsize]{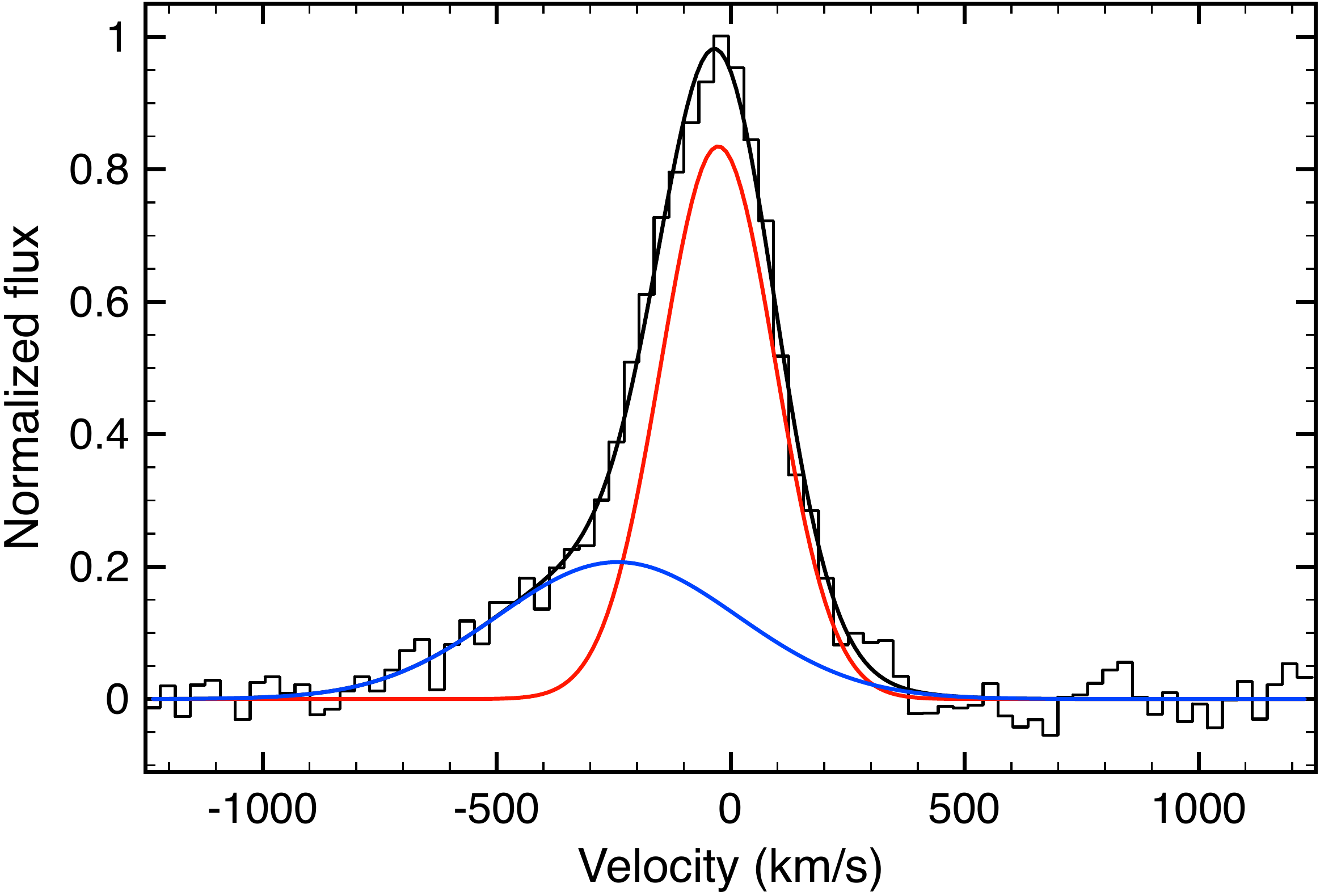}
\caption{Example of a two-component Gaussian fit to the spectrum of IRAS 14348-1447. Details of this spectrum, as well as the spectra of the other ULIRGs, are given in Fig.\,\ref{fig:outflows}.}
\label{fig:spectrum}
\end{figure}

The data were previously presented in PL12 and have a spaxel-size of 0.125$''$, seeing of $\sim$0.6$''$, and dispersion of 2.45\,\AA\,pixel$^{-1}$. The spectral resolution is 6.4$\pm$0.6\,\AA, which corresponds to 90\,$\pm$\,8 \kms\ for the H$_{2}$\,1-0\,S(1) emission line at 2.1218\,$\mu$m. PL12 fitted a single Gaussian to the spatially resolved H$_{2}$\,1-0\,S(1) line to retrieve the global kinematics of the molecular gas. However, in the four ULIRGs that we describe in this paper, the kinematic structure of the hot molecular gas appeared more complex than that. To more accurately derive the gas kinematics, we modified the data analysis routine of PL12 to fit a second Gaussian component to the H$_{2}$ line profiles of these ULIRGs (Fig. \ref{fig:spectrum}). This routine was compiled for the Interactive Data Language (IDL) and uses the MPFIT package for the $\chi^{2}$ minimization \citep{mar09}. We placed no constraints on the fitting parameters and visually inspected all fits to ensure their validity. For each component we created a flux, velocity, and dispersion map, which were boxcar-smoothed spatially by three pixels.

\section{Results}
\label{sec:results}

\begin{figure*}
\centering
\includegraphics[width=\hsize]{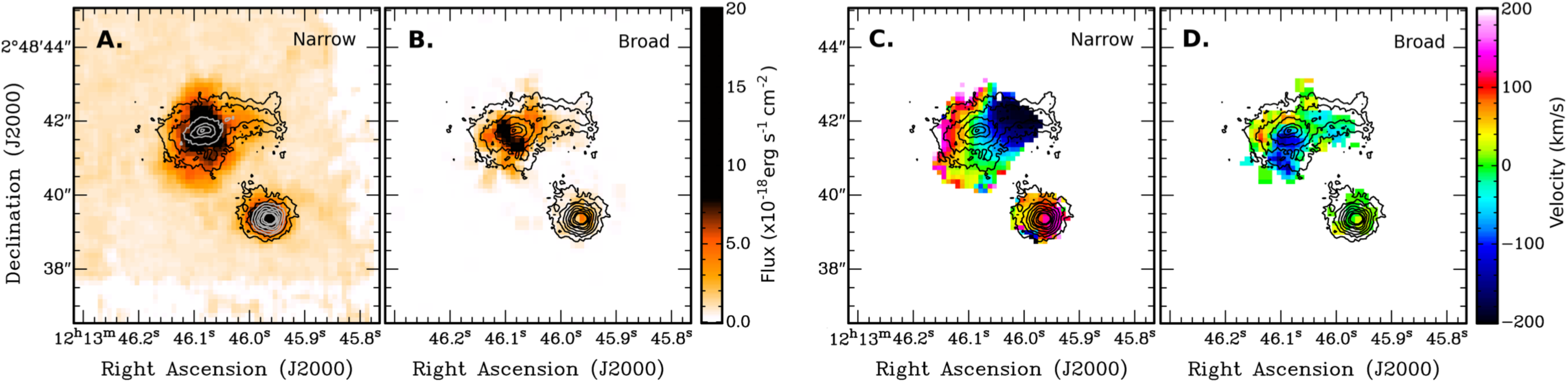}
\caption{Decomposition of the H$_{2}$ 1-0\,S(1) emission in IRAS 12112+0305. Left: Flux of the narrow (A) and broad (B) component of the H$_{2}$ emission. Right: Velocity map of the narrow (C) and broad (D) component of the H$_{2}$ emission in the region where the corresponding flux is $\ge$2\,$\times$\,10$^{-18}$ erg\,s$^{-1}$\,cm$^{-2}$\,spaxel$^{-1}$ for the narrow and $\ge$0.7\,$\times$\,10$^{-18}$ erg\,s$^{-1}$\,cm$^{-2}$\,spaxel$^{-1}$ for the broad component. The dark contours visualize an image taken with the Hubble Space Telescope Near Infrared Camera and Multi-Object Spectrometer (HST/NICMOS) in the combined F220M+F160W+F110W filters \citep{sco00}. HST contours start at $\sim$7$\sigma$, and are drawn at 12, 19, 28, 47, 71, 95, 143, 190$\%$ of the peak intensity of the NE galaxy, which is the galaxy that shows the broad-component emission.}
\label{fig:12112}
\end{figure*}

\begin{figure*}
\centering
\includegraphics[width=\hsize]{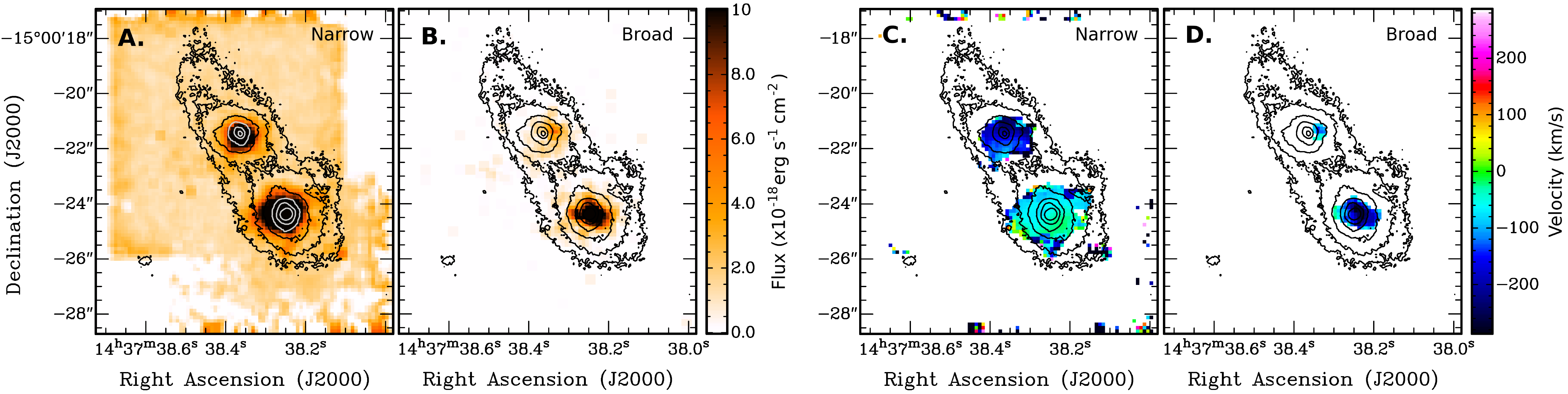}
\caption{Decomposition of the H$_{2}$ 1-0\,S(1) emission in IRAS 14348-1447. Left: Flux of the narrow (A) and broad (B) component of the H$_{2}$ emission. Right: Velocity map of the narrow (C) and broad (D) component of the H$_{2}$ emission in the region where the corresponding flux is $\ge$3\,$\times$\,10$^{-18}$ erg\,s$^{-1}$\,cm$^{-2}$\,spaxel$^{-1}$ for both the narrow and the broad component. The dark contours visualize an image taken with HST/NICMOS in the combined F220M+F160W+F110W filters \citep{sco00}. HST contours start at $\sim$5$\sigma$, and are drawn at 1, 2, 5, 9, 18, 36$\%$ of the peak intensity of the SW galaxy, which is the galaxy that shows the broad-component emission.}
\label{fig:14348}
\end{figure*}

\begin{figure*}
\centering
\includegraphics[width=\hsize]{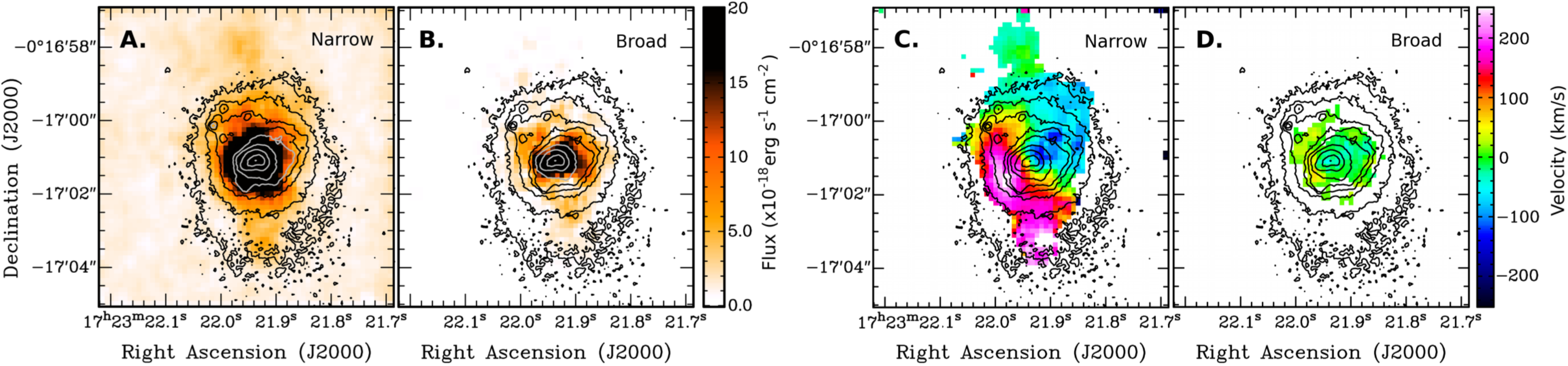}
\caption{Decomposition of the H$_{2}$ 1-0\,S(1) emission in IRAS 17208-0014. Left: Flux of the narrow (A) and broad (B) component of the H$_{2}$ emission. Right: Velocity map of the narrow (C) and broad (D) component of the H$_{2}$ emission in the region where the corresponding flux is $\ge$2.5\,$\times$\,10$^{-18}$ erg\,s$^{-1}$\,cm$^{-2}$\,spaxel$^{-1}$ for the narrow and $\ge$3.5\,$\times$\,10$^{-18}$ erg\,s$^{-1}$\,cm$^{-2}$\,spaxel$^{-1}$ for the broad component. The dark contours visualize an image taken with HST/NICMOS in the combined F220M+F160W+F110W filters \citep{sco00}. HST contours start at $\sim$5$\sigma$, and are drawn at 2.5, 5, 10, 15, 20, 30, 50, 75, 100$\%$ of the peak intensity.}
\label{fig:17208}
\end{figure*}

Figures \ref{fig:12112} $-$ \ref{fig:22491} show the results of our two-component Gaussian fit to the H$_{2}$ 1-0\,S(1) line in IRAS 12112+0305, IRAS 14348-1447, IRAS 17208-0014, and IRAS 22491-1808. In all four ULIRGs, the Gaussian component with the highest peak flux is typically also the narrowest of the two components, with a full width at half the maximum intensity (FWHM) in the range of $\sim$180$-$280 \kms. This component unambiguously traces the gas disk that was previously discussed in these ULIRGs by PL12. We will refer to this primary Gaussian component as the `narrow' component. The secondary Gaussian component reveals emission in the wings of the H$_{2}$ line profile, and is typically broader and less luminous than the primary disk component. We therefore refer to this secondary Gaussian component as the `broad' component, with the caveat that in those spaxels where this secondary component is weak, it may appear narrower than the primary component. This broad-component emission reveals gas features that were not previously found by the single-component Gaussian approach of PL12. In Appendix \ref{sec:app1} we describe the basic properties of the broad-component emission in each ULIRG.

The broad-component emission in the NE galaxy in the IRAS\,12112+0305 pair (Fig. \ref{fig:12112}), the SW galaxy in the IRAS\,14348-1447 pair (Fig. \ref{fig:14348}), and the E galaxy in the IRAS\,22491-1808 pair (Fig.\,\ref{fig:22491}) is shown in more detail in Fig.\,\ref{fig:outflows}. We find that in these systems the broad-component gas is decoupled from the narrow-component disk for two reasons: a) at the location of the most blueshifted emission in the maps of Fig.\,\ref{fig:outflows}, the difference between the peak velocity of the broad and the narrow component in the emission-line profile exceeds the maximum rotational velocity of the narrow-component disk; b) the broad-component gas is blueshifted in the regions where the narrow-component disk is either redshifted or at the systemic velocity. We therefore argue that these three ULIRGs show an outflow of hot molecular H$_{2}$ gas. The emission-line profiles of the H$_{2}$ outflows stretch up to a maximum blueshifted velocity at zero intensity of $\sim$700 \kms\ for IRAS\,12112+0305, $\sim$900 \kms\ for IRAS\,14348-1447, and $\sim$500 \kms\ for IRAS\,22491-1808 (Fig.\,\ref{fig:outflows}). In Sect.\,\ref{sec:parameters}, we will discuss in detail the geometry and mass of the outflows.

As a note of caution, in Appendix \ref{sec:app2} we show that the velocity dispersion of the broad-component emission is significantly lower in IRAS\,22491-1808 than it is in IRAS\,12112+0305 and IRAS\,14348-1447. Although the broad-component emission in IRAS\,22491-1808 appears to be an outflow that is de-coupled from the slowly rotating narrow-component disk (Fig.\,\ref{fig:22491}), the relatively low velocities imply that future observations with JWST or the Atacama Large Millimeter/submillimeter Array (ALMA) are needed to completely rule out internal gas kinematics playing a role in producing the broad-component emission in IRAS\,22491-1808.

The fourth target in our sample, IRAS\,17208-0014, shows broad-component emission that is clearly spatially resolved, but has velocities close to the systemic velocity of the galaxy. The nature of the broad-component emission in IRAS\,17208-0014 is far less clear than it is in IRAS\,12112+0305, 14348-1447, and 22491-1808. We will discuss IRAS\,17208-0014 in detail in Sect.\,\ref{sec:17208}.

\subsection{Physical parameters of the H$_{2}$ outflows} 
\label{sec:parameters}

\begin{figure*}
\centering
\includegraphics[width=\hsize]{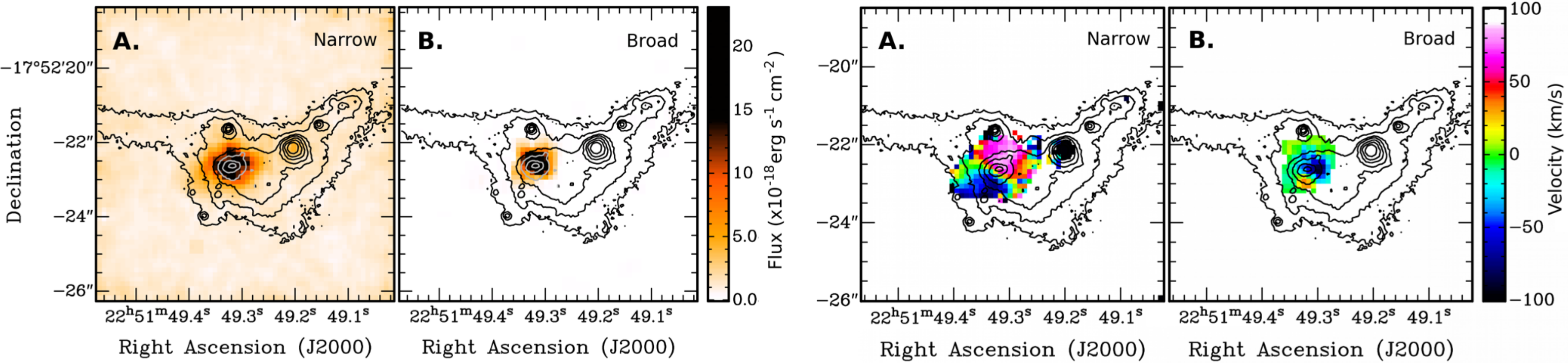}
\caption{Decomposition of the H$_{2}$ 1-0\,S(1) emission in IRAS 22491-1808. Left: Flux of the narrow (A) and broad (B) component of the H$_{2}$ emission. Right: Velocity map of the narrow (C) and broad (D) component of the H$_{2}$ emission in the region where the corresponding flux is $\ge$2.2\,$\times$\,10$^{-18}$ erg\,s$^{-1}$\,cm$^{-2}$\,spaxel$^{-1}$ for the narrow and $\ge$0.4\,$\times$\,10$^{-18}$ erg\,s$^{-1}$\,cm$^{-2}$\,spaxel$^{-1}$ for the broad component. The dark contours visualize an image taken with HST/NICMOS in the combined F220M+F160W+F110W filters \citep{sco00}. HST contours start at $\sim$5$\sigma$, and are drawn at 7, 11, 21, 32, 53, 85, 150$\%$ of the peak intensity of the E galaxy, which is the galaxy that shows the broad-component emission.}
\label{fig:22491}
\end{figure*}

The outflows that we detect in IRAS\,12112+0305, IRAS\,14348-1447, and IRAS\,22491-1808 (Fig.\,\ref{fig:outflows}) are resolved both kinematically and spatially, which makes our SINFONI data ideal for further investigating the physical properties of these outflows.
Table \ref{tab:results} shows the physical parameters that we derive from our measurements. Errors in the Gaussian fitting procedure, as well as the redshift determination, can affect the measurement of the integrated flux of the broad component $F_{\rm broad}$ and associated mass of the outflow. This is particularly true when the velocity difference between the broad and narrow component is small. Therefore, simply summing up the flux of the broad-component in Figs. \ref{fig:12112}, \ref{fig:14348}, and \ref{fig:22491} likely provides a firm upper limit on the total flux of the outflows. A conservative lower limit is given by summing the broad-component flux only in those spaxels where the broad and narrow component are clearly separated kinematically. For this, we only take into account the total broad-component line emission in those spaxels where the difference in central peak velocity between the broad and narrow component, as shown in Figs. \ref{fig:12112}, \ref{fig:14348}, and \ref{fig:22491}, is $|\Delta$v$|$\,$\ge$\,100 km\,s$^{-1}$. In Table\,\ref{tab:results}, $F_{\rm broad}$ is the average value between the upper- and the lower-limit calculation, with corresponding uncertainties reflecting the difference between the two approaches.

The temperature of the H$_{2}$ gas in the outflows can in principle be estimated if multiple rovibrational H$_{2}$ lines are measured \citep[e.g.,][]{dav03}. In the case of IRAS\,12112+0305, IRAS\,14348-1447, and IRAS\,22491-1808 we cannot determine the H$_{2}$ temperature from our SINFONI data, because the H$_{2}$ 1-0\,S(0) ($\lambda_{\rm rest}$\,=\,2.22$\mu$) and 2-1\,S(1) ($\lambda_{\rm rest}$\,=\,2.25$\mu$m) lines are faint and fall in a region of the spectrum that is dominated by strong sky lines. This makes it impossible to derive an accurate flux of the broad-component emission from H$_{2}$ 1-0\,S(0) and H$_{2}$ 2-1\,S(1).

\begin{table*}
\caption{ Physical parameters of the 2.12\,$\mu$m H$_{2}$\,1-0\,S(1) line features. The redshift is indicated by $z$, D$_{\rm L}$ is the luminosity distance (PL12), $F_{\rm broad}$ the flux of the broad component (see Sect.\,\ref{sec:parameters}),
$\Delta$v\,=\,v$_{\rm peak(broad)}$\,-\,v$_{\rm peak(narrow)}$ the velocity difference between peak of the broad and the peak of the narrow component, FWHM the full width at half the maximum intensity of the components in the line-profile, v$_{\rm max(broad)}$\,=\,$\Delta$v\,-\,(FWHM/2) the maximum outflow velocity \citep[e.g.,][]{rup05,arr14}, R$_{\rm broad}$ the total extent of the broad component, M$_{\rm (hot~H_2)}$ the derived mass of the hot molecular gas (Sect.\,\ref{sec:mass}), and \.{M}$_{\rm (total~H_2)}$ the estimated mass outflow rate of the total (hot + cold) molecular gas, assuming a hot-to-cold gas mass ratio of 6\,$\times$\,10$^{-5}$ (Sect.\,\ref{sec:mass}). The measurements of $F_{\rm broad}$ and R$_{\rm broad}$ are derived from the flux and velocity maps of Figs.\,\ref{fig:12112}$-$\ref{fig:22491}, while $\Delta$v, FWHM, and v$_{\rm max(broad)}$ are derived from the H$_{2}$ line profiles shown in Fig\,\ref{fig:outflows}. For additional properties of the narrow-component emission, we refer to PL12 and Appendix \ref{sec:app1}.}        
\label{tab:results}      
\centering          
\begin{tabular}{lcccc}  
 & 12112+0305 (NE) & 14348-1447 (SW)  & 22491-1808 (E) & 17208-0014\\
\hline
$z$  & 0.07310 & 0.08300  & 0.07776 & 0.04281 \\
D$_{\rm L}$ (Mpc) & 336 & 382  & 347& 189 \\
$F_{\rm broad}$ ($\times$10$^{-16}$ erg\,s$^{-1}$\,cm$^{-2}$) & 5.0\,$\pm$\,2.6 & 6.4\,$\pm$\,1.5  & 5.8\,$\pm$\,1.8 & 22\,$\pm$8 \\
$\Delta$v (km\,s$^{-1}$) & -190\,$\pm$\,65 & -215\,$\pm$\,55  & -160\,$\pm$\,75 & $-$  \\
FWHM$_{\rm narrow}$ (km\,s$^{-1}$)$^{\dagger}$ & 265\,$\pm$\,20 & 275\,$\pm$\,15  & 185\,$\pm$\,25 & 255$\pm$20\\
FWHM$_{\rm broad}$ (km\,s$^{-1}$)$^{\dagger}$ & 465\,$\pm$\,50 & 600\,$\pm$\,60  & 310\,$\pm$\,75 & 340$\pm$40 \\
v$_{\rm max(broad)}$  (km\,s$^{-1}$) & -430 & -520  & -320 & $-$ \\
Size$_{\rm broad}$ (kpc) & 1.6 (1.0$^{\prime\prime}$)  & 0.9 (0.5$''$)  &  0.7 (0.4$''$) & 1.9 (2.1$''$) \\
M$_{\rm broad (hot~H_2)}$ (M$_{\odot}$) & (6.8$\pm$3.7)\,$\times$10$^{3}$ & (8.4$\pm$2.2)\,$\times$10$^{3}$  & (5.9$\pm$1.9)\,$\times$10$^{3}$ &  (7.5\,$\pm$\,2.7)\,$\times$10$^{3}$  \\
\.{M}$_{\rm broad (total~H_2)}$ (M$_{\odot}$ yr$^{-1}$)$^{\ddagger}$ & 30 & 85 & 45 & $-$ \\
\end{tabular}
\flushleft 
$^{\dagger}$ Corrected for instrumental broadening of 90 \kms\ (PL12).\\
$^\ddagger$ Based on a hot-to-cold molecular gas mass ratio of $6 \times 10^{-5}$ \citep{emo14,per16}.\\ 
\end{table*}

\subsubsection{Geometry and kinematics of the outflows}
\label{sec:geometry}

The measured size of the outflow region in IRAS\,12112+0305, IRAS\,14348-1447, and IRAS\,22491-1808 ranges from 0.7\,$-$\,1.6\,kpc. These are strict lower limits, given that we do not know, and thus do not corrrect for, the inclination of the outflow. The outflows do not extend across the entire area of the narrow-component disk. In fact, their morphologies suggest that the outflows originate from the nuclear region (see Sect. \ref{sec:app1} for details).

\begin{figure*}
\centering
\includegraphics[width=\hsize]{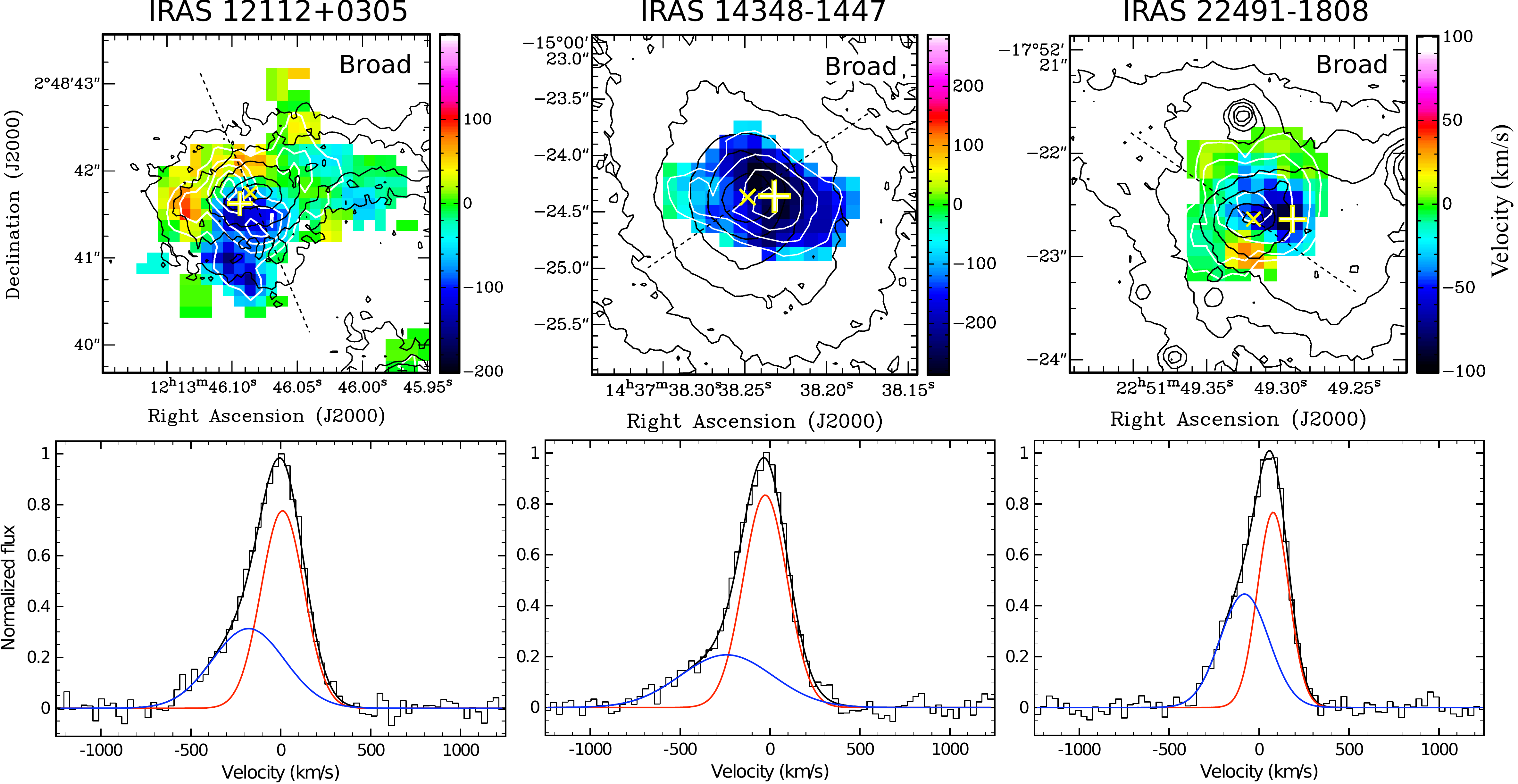}
\caption{Top: Zoom-in of the  H$_{2}$ outflows from Figs. \ref{fig:12112}, \ref{fig:14348}, and \ref{fig:22491}. The H$_{2}$ velocity map of the broad-component emission is shown in color, with overlaid in white contours the H$_{2}$ total-intensity map of the broad component and in black contours the HST/NICMOS image from Figs. \ref{fig:12112}, \ref{fig:14348}, and \ref{fig:22491}. Contour levels of the total-intensity maps of the broad component are: 2, 4, 6, 8, 16 $\times$\,10$^{-18}$ erg\,s$^{-1}$\,cm$^{-2}$\,spaxel$^{-1}$ (IRAS\,12112+0305); 5, 9, 13, 17 $\times$\,10$^{-18}$ erg\,s$^{-1}$\,cm$^{-2}$\,spaxel$^{-1}$ (IRAS\,14348-1447); 1, 3, 9, 27 $\times$\,10$^{-18}$ erg\,s$^{-1}$\,cm$^{-2}$\,spaxel$^{-1}$ (IRAS\,22491-1808). The small cross ($\times$) marks the peak of the IR-emission in the NICMOS image, which is presumably the center of the galaxy. The large cross (+) marks the location where the outflow reaches a maximum velocity at zero intensity, and against which we extracted the H$_{2}$ spectra shown below. The dashed line represents the direction of the minor axis, as we estimated from the kinematic of the narrow-component disk and the HST/NICMOS imaging (see Appendix \ref{sec:app1}). Bottom: H$_{2}$ 1-0\,S(1) spectra extracted from a region of five pixels (three pixels both E to W and N to S, as marked with the `+' sign above). The black line shows a double Gaussian fit, consisting of a narrow (red) and broad (blue) component.}
\label{fig:outflows}
\end{figure*}

Figure \ref{fig:outflows} suggests that the outflows in IRAS\,12112+0305 and IRAS\,22491-1808 appear to be aligned somewhat closer to the minor than to the major axis of the galaxies, although not to such an extent that we draw any firm conclusions from this. For IRAS\,14348-1447, the galaxy appears to be oriented close to face-on, while the high radial velocity and small spatial extent of the outflow likely indicate that the gas is flowing mostly along our line of sight. For all three ULIRGs, the outflow velocity reaches a maximum at roughly 0.5 kpc distance from the core. These properties are similar to those of the conical, nuclear H$_{2}$ outflow that was first imaged with near-IR integral-field spectroscopy in QSO F08572+3915 \citep{rup13b}.

The velocity FWHM of the broad-component emission in the outflows is roughly twice that of the narrow-component emission in the disk. In part, this may be related to a backflow of gas, or to the fact that the opening angle of the outflowing gas causes a superposition of radial velocities along our line of sight. However, in particular in the case of IRAS\,12112+0305, the conical outflow appears to have a modest opening angle of $\sim$40$^{\circ}$ (Sect.\,\ref{sec:app1}). It is therefore more likely that the turbulence of the H$_{2}$  gas in the outflow is higher than that in the narrow-component disk. With a 2$\times$ larger FWHM, the turbulent kinetic energy of the gas in the outflow, $E^{\rm turb}_{\rm kin}$\,$=$\,${\frac{3}{2}}$\,$\cdot$\,${\rm M}$\,$\sigma^{2}$ ($\sigma$\,=\,FWHM/2.35), is up to a factor of four higher than that of the gas in the disk.

\subsubsection{H$_{2}$ mass of the outflows}
\label{sec:mass}

To estimate the mass of the hot molecular gas that is locked up in the broad-component emission, we follow \citet{sco82} and \citet{maz13}:
\begin{equation}
{\rm M}_{\rm H_2} \simeq 5.1 \times 10^{13} \left(\frac{{\rm D}_{\rm L}}{{\rm Mpc}}\right)^2 \left(\frac{F_{\rm 1-0S(1)}}{{\rm erg\,s^{-1}\,cm^{-2}}}\right) 10^{(0.4\,\cdot\,{\rm A}_{2.2})},
\label{eq:mass}
\end{equation}
where M$_{\rm H_2}$ is the hot molecular H$_{2}$ mass in M$_{\odot}$, D$_{\rm L}$ is the luminosity distance to the galaxy, and A$_{\rm 2.2\mu m}$ is the extinction at 2.2$\mu$m. In this case, we substitute $F_{\rm 1-0S(1)}$ for the flux of the broad component ($F_{\rm broad}$ in Table\,\ref{tab:results}). This approach assumes thermalized gas conditions and T\,=\,2000\,K, with a population fraction in the ($\nu,J$) = (1,3) level of $f_{\rm (1,3)} = 0.0122$. If a fraction of the H$_{2}$ gas was sub-thermally excited, $f_{1,3}$ would be lower and Equation \ref{eq:mass} would thus give a lower limit to the H$_{2}$ mass. Following \citet{piq13}, A$_{2.2\mu \rm m} \approx 0.1$\,$\times$\,A$_{\rm V}$, with A$_{\rm V}$ the visual extinction, based on the extinction law described by \citet{cal00} \citep[see also, e.g.,][]{rie85,fit99,ind05}. Because the broad-component H$_{2}$ emission is resolved on scales of a few kpc, we use A$_{\rm V, R(eff)}$ from \citet{piq13}, which is the visual extinction within the effective radius of the H$\alpha$ emission \citep{arr12}. We note, however, that this implicitly assumes that the extinction in the outflow is similar to that inferred from the overall (narrow+broad component) H$_{2}$ gas in this region.

Based on Equation \ref{eq:mass}, the broad-component emission in IRAS\,12112+0305, IRAS\,14348-1447, and IRAS\,22491-1808 represents outflows of hot molecular gas with masses of M$_{\rm H2(hot)}$ $\sim$ 6$-$8\,$\times$10$^{3}$ M$_{\odot}$ (Table\,\ref{tab:results}). The total mass of the outflowing molecular gas consists of a hot and cold component. Recently, we derived hot-to-cold molecular gas mass ratios of 6$-$7\,$\times$\,10$^{-5}$ for H$_{2}$ outflows in two nearby luminous infra-red galaxies (LIRGs; 10$^{11}$\,$\le$\,$L_{\rm IR}$\,$<$\,10$^{12}$ $L_{\odot}$), by comparing SINFONI and ALMA data \citep{emo14,per16}. The most prominent of these two outflows, in the LIRG NGC\,3256, has a hot molecular gas mass of $\sim$1200 M$_{\odot}$, which is roughly a factor of six lower than what we observe in our ULIRGs. If we assume a similar hot-to-cold molecular gas mass ratio of 6\,$\times$\,10$^{-5}$, then the  total molecular gas masses involved in the ULIRG outflows are of the order of M$_{\rm H2}$ $\sim$ 1\,$\times$\,10$^{8}$ M$_{\odot}$. This corresponds to mass outflow rates of \.{M}$_{\rm H_2}$ = \(\frac{{\rm M}_{\rm H2(total)}\,\cdot\,{\rm v}_{\rm max}\,{\rm sin(}i{\rm )}}{{\rm R}_{\rm outflow}\,{\rm cos(}i{\rm )}}\) $\approx$ 30$-$85 tan({\sl i}) M$_{\odot}$ yr$^{-1}$, where R$_{\rm outflow}$ is the size of the outflow region, v$_{\rm max}$ the maximum outflow velocity (Table\,\ref{tab:results}), and {\sl i} the line-of-sight inclination.\footnote{Because we do not know the inclination of these systems, we will ignore the inclination effects in the remainder of this paper.} Upcoming CO observations in the millimeter regime will be able to confirm these estimates.

\begin{figure*}
\centering
\includegraphics[width=\hsize]{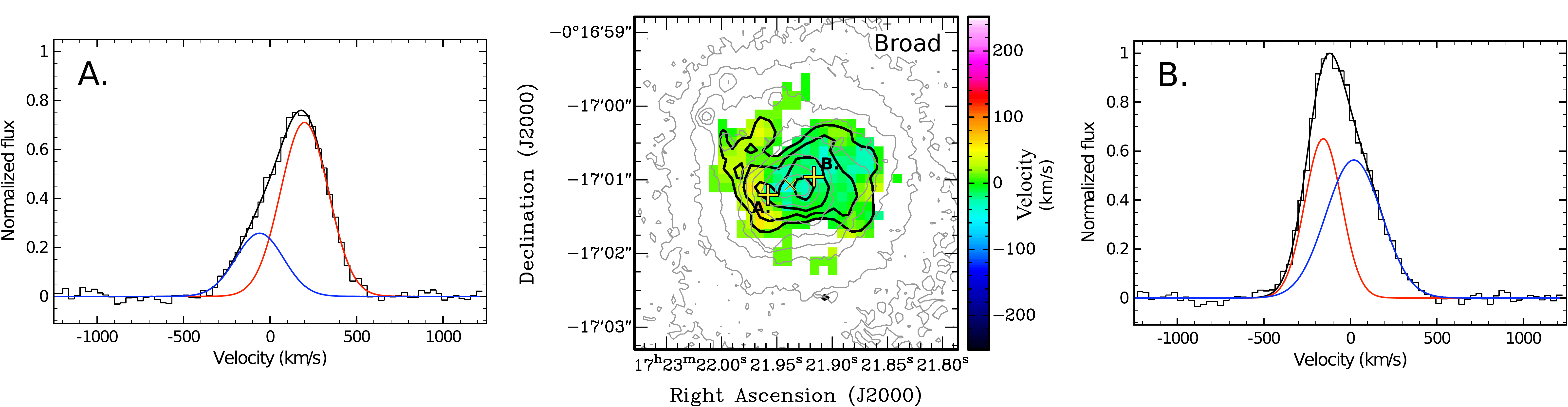}
\caption{Middle: Zoom-in of the broad-component H$_{2}$ emission in IRAS\,17208 from Fig. \ref{fig:17208}. The H$_{2}$ velocity map of the broad-component emission is shown in color, with overlaid in thick black contours the H$_{2}$ total-intensity map of the broad component and in thin gray contours the HST/NICMOS image from Fig. \ref{fig:17208}. Contour levels of the total-intensity maps of the broad component are: 6, 12, 24, 48 $\times$\,10$^{-18}$ erg\,s$^{-1}$\,cm$^{-2}$\,spaxel$^{-1}$. The small yellow/black cross ($\times$) marks the center of the galaxy from the HST/NICMOS image, while the large black/yellow crosses (+) mark the location against which we extracted the H$_{2}$ spectra A and B. Left/right: H$_{2}$ 1-0\,S(1) spectra extracted from a region of five pixels (three pixels both E to W and N to S, as marked with the `+' signs in the middle panel). The black line shows a double Gaussian fit, consisting of a narrow (red) and broad (blue) component.}
\label{fig:outflow17208}
\end{figure*}

\subsection{IRAS\,17208-0014}
\label{sec:17208}

IRAS\,17208-0014 shows broad-component emission that is clearly spatially resolved, but shows little kinematic structure. The overall H$_{2}$ kinematics in IRAS\,17208-0014 appear to follow the CO(2-1) kinematics of cold molecular gas discussed in detail by \citet{gar15}. We argue that it is unlikely that the broad-component H$_{2}$ emission represents an outflow similar to what we see in IRAS\,12112+0305, IRAS\,14348-1447, and IRAS\,22491-1808. The reason is that the broad-component gas in IRAS\,17208-0014 has a velocity that is close to the systemic velocity across the narrow-component disk. Moreover, the broad-component emission is spread across a large fraction of the narrow-component disk. Therefore, if this broad-component emission represents an outflow, it would have to be gentle and widespread, reminiscent of a galactic wind with very low radial velocity. It is interesting to note that a widespread outflow of neutral gas has been reported in Na\,D absorption across several kpc \citep{mar06,rup13a}. To the contrary, there is no observational evidence for a similar widespread outflow in the ionized gas phase (e.g., \citealt{arr03}, \citealt{wes12}, \citealt{rup13a}; see Sect. \ref{sec:app_17208}).

However, it is good to keep in mind that the broad-component emission is separated from the narrow-component emission as a result of our Gaussian fitting procedure. In fact, in the central kpc-scale region, the morphology of the broad-component emission seems to resemble the structure seen with the Near Infrared Camera and Multi-Object Spectrometer (NICMOS) on the Hubble Space Telescope (HST), while fainter broad-component emission further out shows tentative indications for a spiral-like morphology (Figs.\,\ref{fig:17208} and \ref{fig:outflow17208}). It is therefore possible that the broad-component emission is merely a low-velocity wing to the narrow-component gas, perhaps representing internal gas dynamics or gas falling into the center of the galaxy. Alternatively, the broad-component H$_{2}$ emission could present low-velocity gas outside the galaxy disk, which was previously deposited through a galactic-scale outflow and possibly mixed with low-angular-momentum gas from the halo \citep[see, e.g.,][]{fra06}. Similar low-velocity gas has been seen outside the disks of nearby galaxies in neutral hydrogen gas (e.g., \citealt{boo05}, \citealt{oos07}), and bears resemblance to the low-velocity Na\,D component seen in the LIRG IRAS\,F11506-3851 by \citet{caz14}. Finally, \citet{med14} argue that the core of IRAS\,17208-0014 consists of two close nuclei separated by $\sim$200 pc. It is interesting to speculate that the narrow- and broad-component emission result from two superposed gas disks. Based on Fig.\,\ref{fig:17208}, the dominant narrow-component disk would be substantially inclined. To the contrary, the lack of line-of-sight kinematics among the broad-component emission indicates that this would represent a face-on disk with a size of at least $\sim$2 kpc and possible spiral-type morphology. While the nature of the broad-component emission in IRAS 17208-0014 remains unclear, detailed observations with ALMA or JWST will be able to discern between the different scenarios.

\subsubsection{Upper limit on the H$_{2}$ outflow in IRAS\,17208-0014}
\label{sec:appA_17208}

\begin{figure}
\centering
\includegraphics[width=0.80\hsize]{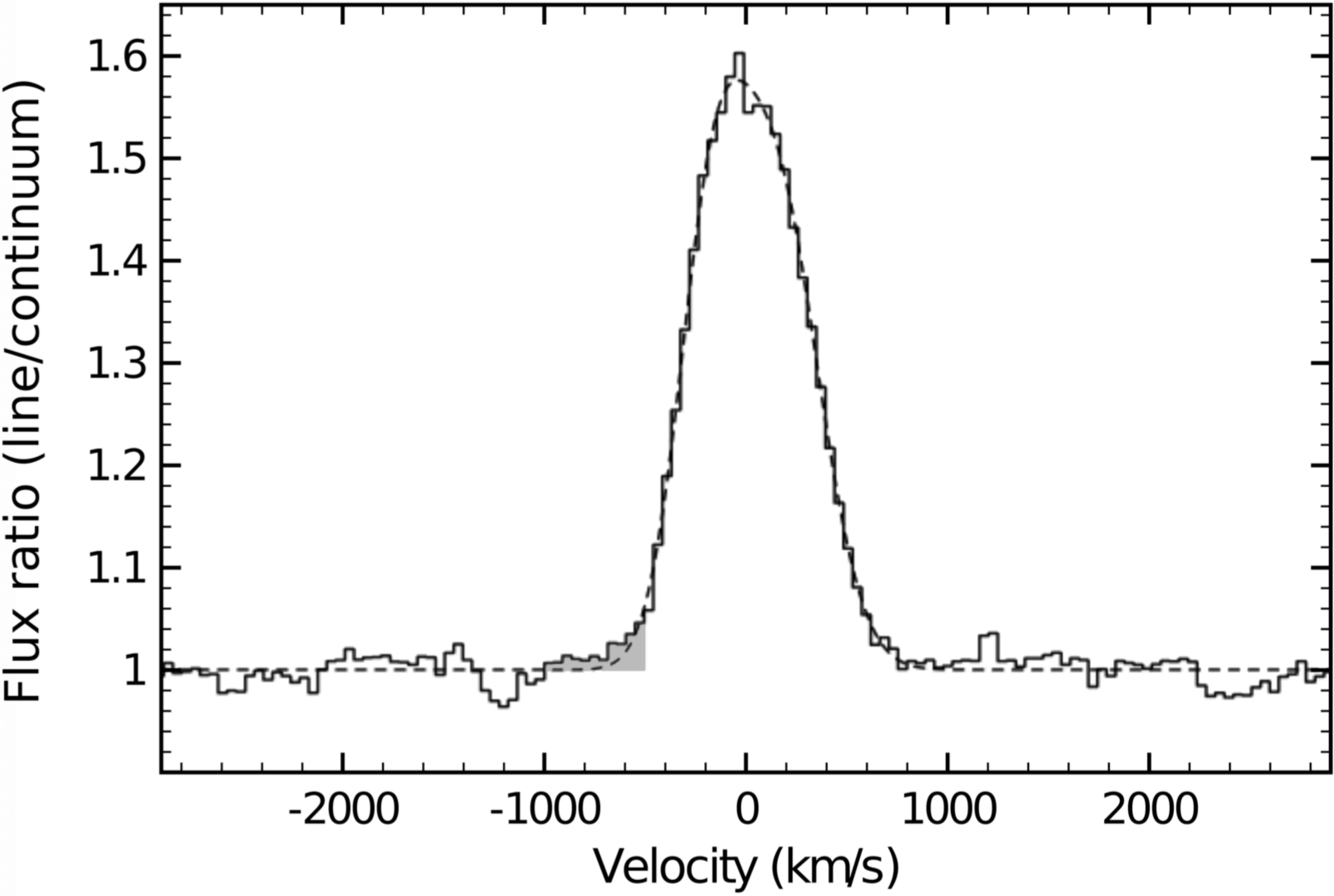}
\caption{Spectrum of the H$_{2}$ 2.1218 $\mu$m line in IRAS\,17208-0014 centered one spaxel (0.125$^{\prime\prime}$) south of the core, where \citet{gar15} found an outflow in CO(2-1). The spectrum was extracted against a circular aperture with an area of 13 pixels, and therefore includes the strong nuclear near-IR continuum emission. Plotted on the y-axis is the ratio of the emission-line flux and the continuum flux. The dashed line shows the best-fit model of the narrow-component disk. No convincing outflow is seen in H$_{2}$. The emission in the gray shaded area, which stretches from -1000\,$-$\,-500 \kms and is affected by noise and continuum features, comprises an H$_{2}$ mass of $\sim$110 M$_{\odot}$. This is a factor of four below the firm upper limit that we derive for an H$_{2}$ outflow in IRAS\,17208-0014 (see text for details).}
\label{fig:spec17208outfl}
\end{figure}

\citet{gar15} also discovered a faint, blueshifted CO(2-1) outflow in IRAS\,17208-0014. Figure \ref{fig:spec17208outfl} shows the H$_{2}$ 1-0\,S(1) spectrum extracted across a circular aperture with an area of 13 spaxels at the location where the CO outflow was found. We do not detect an H$_{2}$ counterpart to this CO outflow in our SINFONI data. Figure \ref{fig:spec17208outfl} shows that any potential outflow is buried well within the noise, and affected by features within the near-IR continuum. The CO outflow stretches roughly from -500 to -700 \kms, which is outside the line profile of the narrow-component H$_{2}$\,1-0\,S(1) emission that we observe with SINFONI (Fig.\,\ref{fig:outflow17208}). Our SINFONI data show an rms noise level of $\sim$8\,$\times$\,10$^{-16}$ erg\,s$^{-1}$\,cm$^{-2}$\,$\mu$m$^{-1}$ per 0.125$''$ spaxel and per 2.45\AA\ (35 \kms) resolution element in the dispersion direction. We set a conservative upper limit on the integrated line flux of a spatially unresolved H$_{2}$\,1-0\,S(1) outflow across 200 \kms\ by assuming a 10$\sigma$ detection when binning the data across 200 \kms\ (six pixels) in dispersion direction and across the 0.6$''$ (18 pixels) seeing-disk in spatial directions. This is equivalent to 1$\sigma$ per raw pixel in the unbinned data for a signal distributed across the seeing-disk and across 200 \kms. Following Sect.\,\ref{sec:mass}, this results in an upper limit on the outflow in hot molecular H$_{2}$ gas of M$_{\rm H2-hot}$ $\la$400 M$_{\odot}$. For comparison, the H$_{2}$ emission from the shaded region in Fig. \ref{fig:spec17208outfl} only comprises M$_{\rm H2-hot}$ $\la$110 M$_{\odot}$. Given the underlying near-IR continuum emission, we therefore argue that M$_{\rm H2-hot}$ $\la$400 M$_{\odot}$ is a conservative upper limit.

With our assumed hot-to-cold molecular gas mass ratio of 6\,$\times$\,10$^{-5}$ (Sect.\,\ref{sec:mass}), the corresponding upper mass limit on the total (hot+cold) molecular outflow is M$_{\rm outflow}$ $\la$ 6.7$\times$10$^{6}$ M$_{\odot}$, which is about a factor of seven lower  than the CO outflow of M$_{\rm outflow}$ $\sim$ 4.6$\times$10$^{7}$ M$_{\odot}$ observed by \citet{gar15}. To bring both observations into agreement, we would have to put an upper limit on the hot-to-cold molecular gas mass ratio of $\la$9\,$\times$\,10$^{-6}$. This is substantially lower than the hot-to-cold molecular gas mass ratios of 6$-$7\,$\times$\,10$^{-5}$ measured directly in the outflows of two nearby LIRGs with SINFONI and ALMA \citep{emo14,per16}. Nevertheless, it is within the 10$^{-7}$\,$-$\,10$^{-5}$ range of generic values previously found by \citet{dal05} in samples of starburst galaxies and AGN. Therefore, our SINFONI data are not sensitive enough to either verify or rule out the existence of a hot H$_{2}$ counterpart to the CO outflow found by \citet{gar15}.

\section{Discussion}

We discovered massive outflows of hot molecular H$_{2}$ gas from three of four nearby ULIRGs (75$\%$), namely IRAS\,12112 (NE), IRAS\,14348 (SW), and IRAS\,22491 (E). These three systems are all in a pre-coalescent stage, and for all three we detect an outflow in only one of the two merging galaxies. We therefore detect a hot molecular outflow in three of the seven individual galaxies in our sample (43$\%$). However, we stress that our SINFONI data are only sensitive to tracing molecular gas outflows of M$_{\rm H2\,(hot+cold)}$\,$\sim$\,10$^{8}$ M$_{\odot}$ with outflow velocities v$_{\rm max}$ $\sim$ several 100 \kms\ (Table \ref{tab:results}). For example, we lack the sensitivity to trace the hot H$_{2}$ counterpart to the molecular outflow that was previously discovered in CO in the fourth ULIRG in our sample, IRAS\,17208-0014 \citep[][see Sect.\,\ref{sec:appA_17208}]{gar15}. Sensitive near-IR integral-field spectroscopy of larger samples of ULIRGs is needed to derive meaningful conclusions about the detection rate of hot molecular H$_{2}$ outflows in ULIRGs.

\subsection{Link to starburst/AGN activity}
\label{sec:link}

IRAS\,12112+0305, IRAS\,14348-1447, IRAS\,17208-0014, and IRAS\,22491-1808 all have a prominent starburst component. Their nuclei are all radio-quiet, and have been classified as either low ionization nuclear emission-line regions (LINERs) or starburst/H\,{\sc II} regions \citep[e.g.,][]{vei99,fra03,nar10}. From high-resolution IR spectroscopy, \citet{alo16} show that IRAS\,14348-1447 and IRAS\,17208-0014 contain a faint AGN at mid-IR wavelengths. 

However, X-ray observations classify the SW nucleus of IRAS\,14348-1447 as a Compton-thick AGN candidate \citep{iwa11}. Previous work revealed that ionized and molecular gas outflows are on average faster and more massive in the presence of an AGN \citep[][see also models by \citealt{nar08}]{spo13, arr14,cic14,hil14}. The fact that IRAS\,14348-1447 (SW) shows the fastest mass outflow among our sample sources is in agreement with this previous work, and suggests that the H$_{2}$ outflow in IRAS\,14348-1447 is driven by the Compton-thick AGN.

The energetics in IRAS\,14348-1447 are also consistent with this scenario. \citet{iwa11} estimate that the X-ray luminosity of the SW galaxy of IRAS\,14348-1447 is $L_{\rm X\,(0.5-10 keV)}$ $\sim$ 5.4\,$\times$\,10$^{41}$ erg\,s$^{-1}$. Following \citet{elv94} and \citet{ho08}, the expected bolometric output of the AGN is roughly 20$\times$ $L_{\rm X\,(0.5-10 keV)}$, or $L_{\rm bol}$\,$\times$\,1.1$\times$10$^{43}$ erg\,s$^{-1}$. The H$_{2}$ outflow has a bulk flow of $\sim$200 \kms, which means that it has taken roughly 5\,$\times$\,10$^{6}$ yr for the flow to reach $\sim$1 kpc distance (Table \ref{tab:results}). In 5\,$\times$\,10$^{6}$ yr, the estimated energy output of the Compthon thick AGN is E$_{\rm AGN}$\,$\sim$\,1.7\,$\times$\,10$^{57}$ erg. This is likely a lower limit, given that the 0.5$-$10 keV emission could be heavily absorbed. The total (bulk\,+\,turbulent) kinetic energy of the outflow is \(E_{\rm tot} = E^{\rm turb}_{\rm kin} + E^{\rm bulk}_{\rm kin} = {\frac{3}{2}} \cdot {\rm M}\sigma^{2} + {\frac{1}{2}} \cdot {\rm M}{\rm [v/sin({\it i}_{\rm outfl})]}^{2}\) = 3.4\,$\pm$0.9\,$\times$\,10$^{56}$ erg (following Table \ref{tab:results}). This is a strict lower limit, given that the inclination $i_{\rm outfl}$ is likely to be large. Nevertheless, the energy output of the Compton-thick AGN appears to be sufficient to drive the outflow in IRAS\,14348-1447.

\subsection{Effect on galaxy evolution}

Based on spectral energy distributions from the ultra-violet to far-infrared, \citet{cun10} derive stellar masses of 2.2, 13.5, and 4.2 $\times$10$^{10}$ M$_{\odot}$ for IRAS\,12112+0305, IRAS\,17208-0014, and IRAS\,22491-1808, respectively. Assuming that this stellar mass is distributed equally between the merging galaxies in each ULIRG, the escape velocity at 2 kpc distance from the galaxy ranges from $\sim$220 \kms\ for IRAS\,12112+0305 (NE) to $\sim$300 \kms\ for IRAS\,22491-1808 (E). \citet{arr14} instead derive dynamical masses of 4.6, 7.4, 2.1, and 3.4 $\times$10$^{10}$ M$_{\odot}$ for IRAS\,12112+0305, IRAS\,14348-1447, IRAS\,17208-0014, and IRAS\,22491-1808, respectively. Following Equation 7 of \citet{arr14}, assuming r\,=\,2\,kpc and an isothermal sphere with truncation radio r$_{\rm max}$\,=\,20 kpc, this results in v$_{\rm esc}$\,$\sim$\,465, 590, and 400 for the merging ULIRGs IRAS\,12112+0305, 14348, and 22491-1808, respectively. Therefore, while a small fraction of the outflowing molecular gas may escape the ULIRGs, the bulk of the outflowing molecular gas in IRAS\,12112-0305 and IRAS\,14348-1447, and possibly all of it in IRAS\,22491-1808, will remain bound to the merging system. The outflows that we observe will therefore mainly re-distribute the gas within the ULIRGs, which may contribute to the dynamical processes that delay the formation of stars. However, the molecular gas itself will remain available within the system to fuel future star formation.

Moreover, \citet{eva02} and \citet{gar09} derive star formation rates of the order of 200\,$-$\,400 M$_{\odot}$\,yr$^{-1}$ for IRAS\,12112-0305 and IRAS\,14348-1447, based on infrared, radio, and H$\alpha$ observations. Our estimated molecular outflow rates of \.{M}$_{\rm H_2 (total)}$ $\sim$ 30\,$-$\,85 M$_{\odot}$\,yr$^{-1}$ are at least a factor of a few lower than the star formation rates. This suggests that the effects that the molecular outflow have on delaying star formation are rather modest.

\subsection{Multi-phase gas outflows}

Multi-phase gas outflows, down to the lowest temperatures of molecular gas, can be an important aspect in the evolution of ULIRGs. A demonstration of this are the detailed studies done on the nearby quasar-type ULIRG MRK\,231 \citep[e.g.,][]{rup11,aal15,ala15,fer15,mor16,vei16}. Such multi-phase gas outflows are likely to have cold, dense clouds entrained and accelerated by a flow of warm and hot gas \citep[e.g.,][]{wag12,zub14}. A clumpy outflow of cold molecular gas was recently observed with ALMA in the LIRG IRAS F11506-3851, where CO-emitting clouds with sizes between 60 and 150 pc are part of a multi-phase outflow \citep{per16}.

The best studied phase of outflows is the ionized gas phase, as traced by optical emission lines. While a detailed discussion of ionized outflows is beyond the scope of this paper, it is interesting to note that among a large sample of nearby LIRGs and ULIRGs, \citet{arr14} found that outflows are nearly ubiquitous in (U)LIRGs. The outflow velocities that they observe are similar to what we detect in H$_{2}$ in our small sample of ULIRGs, while also the ionized gas outflows do not reach the escape velocity in all but the least massive systems \citep{arr14}.

Considering  the molecular gas, the nearby LIRG NGC\,3256 was the first galaxy where we imaged a bi-conical molecular outflow both in the hot and cold molecular phase at $\sim$100\,pc resolution \citep[][see also \citealt{sak14}]{emo14}. The hot and cold components share a similar morphology and kinematics, which suggests that the hot and cold phase trace the same gas features. In NGC\,3256 we derived a hot-to-cold molecular gas mass ratio of $\sim$6\,$\times$\,10$^{-5}$, which is similar to that found in another nearby LIRG \citep{per16}. The total molecular mass outflow rate of at least \.{M}$_{\rm H_2}$\,$\ga$\,20 M$_{\odot}$\,yr$^{-1}$ in the LIRG NGC\,3256 is similar to that of the H$_{2}$ outflows that we identified in the ULIRGs studied in this paper.

On the other hand, \citet{cic14} studied molecular outflows through observations of CO in a sample of nearby ULIRGs. They find the presence of molecular outflows with mass outflow rates of several 100 M$_{\odot}$\,yr$^{-1}$, which are similar or larger than the star formation rates in these systems. For the ULIRGs that we study in this paper, we derive outflow rates of molecular gas that are a factor of a few lower than the star formation rates (Sect.\,\ref{sec:link}). Therefore, our H$_{2}$ results appear at odds with the CO results from \citet{cic14}. Of course, the H$_{2}$ 2.12$\mu$m line only traces a small fraction of the total molecular gas mass, and its strength is subject to the local gas temperature and excitation \citep[see, e.g.,][]{rig02}. Complementary CO or mid-IR H$_{2}$ observations of our sample sources are needed to further investigate the hot-to-cold molecular gas mass ratio.

Some of the most spectacular molecular gas outflows in ULIRGs have been detected with the Herschel Space Observatory in the OH line \citep[e.g.,][]{stu11,spo13,vei13,gon17}. The outflow velocities that we observe in the H$_{2}$ line with SINFONI are within the range of the OH outflows detected with Herschel. However, the mass outflow rates derived from OH measurements by \citet{stu11} and \citet{gon17} are nearly ubiquitously \.{M}$_{\rm H_2}$\,$\ga$\,100 M$_{\odot}$\,yr$^{-1}$. Our SINFONI data suggest also here that either molecular outflow rates in nearby ULIRGs are much more moderate than predicted based on OH studies, or that our assumed hot-to-cold gas mass ratio for H$_{2}$ outflows in ULIRGs is too large. Alternatively, the OH studies may trace faster outflows that originate closer to the galaxy core, with higher v$_{\rm max}$ and smaller R$_{\rm outflow}$. \citet{gon17} indeed found evidence for higher maximum outflow velocities seen in OH compared to CO in several nearby ULIRGs. According to \citet{gon17}, this could suggest that OH and CO trace outflows in different regions of the galaxy. After all, they argue that as the outflowing gas expands and breaks up into clumps, it will ``cover a decreasing fraction of the far-IR continuum''. Unlike the CO, the OH molecules are radiatively excited, and hence ``produce lower absorption and emission'' the further out the gas flows \citep{gon17}. This scenario may also explain the fact that the molecular mass outflow rates that we derive from H$_{2}$ appear on average lower than those based on OH studies.

Our results suggest that consensus has yet to be reached on how effective molecular gas outflows are in influencing the star formation properties of nearby ULIRGs. We argue that detailed 3D-imaging of the different gas phases is a valuable tool to help determine the evolutionary role of multi-phase gas ouflow in the evolution of these IR-luminous galaxies.

\section{Conclusions}

We presented the detection and characterization of massive outflows of hot molecular gas in three out of four nearby ULIRGs. We imaged the molecular outflows at sub-kpc resolution with VLT-SINFONI, using the H$_{2}$ 1-0\,S(1) 2.12$\mu$m line. We derive the following conclusions:\\
\vspace{-2mm}\\
$\bullet$ The H$_{2}$ outflows spread from the nuclear region out to scales of 0.7$-$1.6 kpc;\\
\vspace{-2mm}\\
$\bullet$ The H$_{2}$ outflows contain a hot molecular gas mass of M$_{\rm H2(hot)}$\,$\sim$6$-$8\,$\times$\,10$^{3}$ M$_{\odot}$. Assuming a hot-to-cold H$_{2}$ ratio of 6\,$\times$\,10$^{-5}$, as previously found in outflows from LIRGs, this corresponds to total molecular gas masses of M$_{\rm H2}$\,$\sim$\,10$^{8}$ M$_{\odot}$;\\
\vspace{-2mm}\\
$\bullet$ The corresponding total mass outflow rates of molecular gas are \.M$_{\rm H2}$\,$\sim$\,30$-$85 M$_{\odot}$\,yr$^{-1}$, which is a factor of five to ten lower than the star formation rates in these systems;\\
\vspace{-2mm}\\
$\bullet$ The typical outflow velocities range from 300\,$-$\,500 \kms, which suggests that a large fraction of outflowing gas remains gravitationally bound and is merely re-distributed within the ULIRGs;\\
\vspace{-2mm}\\
$\bullet$ The fastest outflow is associated with the Compton-thick AGN candidate IRAS\,14348-1447, with a velocity-wing in the emission-line profile that stretches out to $\sim$900 \kms\ at zero intensity.\\
\vspace{-2mm}\\
We do not find convincing evidence that the molecular outflows that we discovered in the H$_{2}$ 2.12$\mu$m line in three of the nearest ULIRGs dramatically change the evolution of these systems. Future studies with JWST and ALMA will provide the sensitivity to further investigate how common and how destructive molecular outflows are in ULIRGs throughout the Universe.

\begin{acknowledgements}
We thank the anonymous referee for valuable feedback that helped improve the paper. The research leading to these results was funded by the Spanish Ministerio de Econom\'{i}a y Competitividad (MINECO) under grants AYA2012-32295, ESP2015-68964-P, AYA2016-76682-C3-2-P, and AYA2015-54346-C2-1-P. Based on observations collected at the European Organisation for Astronomical Research in the Southern Hemisphere, Chile, progs. 077.B-0151A and 081.B-0042A. Based on observations made with the NASA/ESA Hubble Space Telescope, and obtained from the Hubble Legacy Archive, which is a collaboration between the Space Telescope Science Institute (STScI/NASA), the Space Telescope European Coordinating Facility (ST-ECF/ESA), and the Canadian Astronomy Data Centre (CADC/NRC/CSA). The National Radio Astronomy Observatory is a facility of the National Science Foundation operated under cooperative agreement by Associated Universities, Inc.
\end{acknowledgements}

\begin{appendix} 

\section{Individual sources}
\label{sec:app1}

In the following sections we provide detailed information regarding the broad component H$_{2}$ emission in IRAS\,12112+0305, IRAS\,14348-1447, IRAS\,17208-0014, and IRAS\,22491-1808.

\subsection{IRAS\,12112+0305} 

\begin{figure}[b!]
\centering
\includegraphics[width=0.80\hsize]{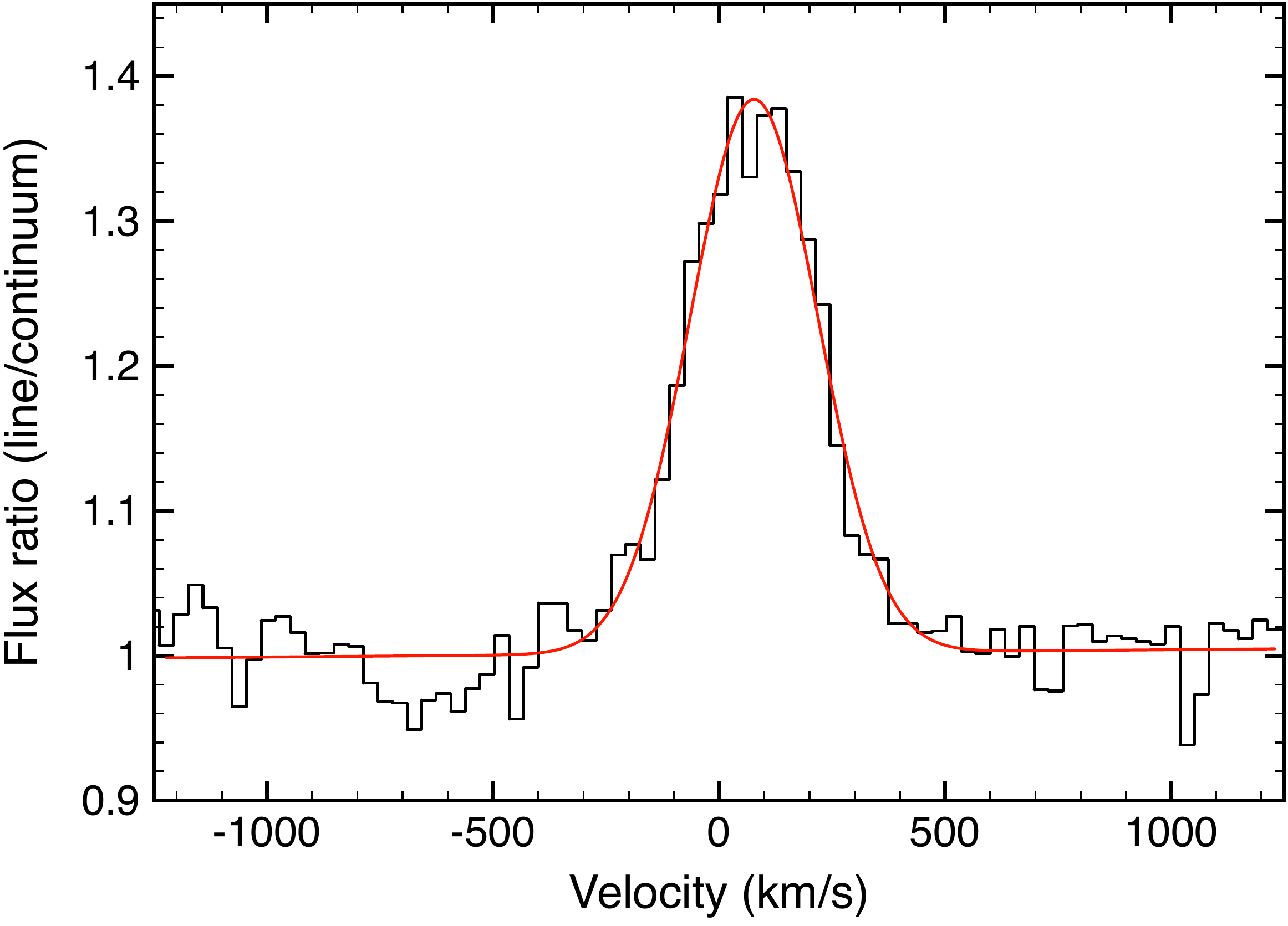}
\caption{Spectrum of the H$_{2}$ 1-0\,S(1) line in the SW galaxy of the IRAS\,12112+0305 galaxy-pair, that is, the galaxy that does not host the prominent outflow. Plotted on the y-axis is the ratio of the emission-line flux and the continuum flux. The spectrum is extracted from a region of five pixels (three pixels both E to W and N to S), centered on the peak of the continuum emission. The red line shows a single Gaussian function, which accurately fits the H$_{2}$ line. This plot shows that the weak broad-component emission from Fig. \ref{fig:12112} is merely a result of our uncertainty in the Gaussian fitting procedure due to the underlying strong continuum emission of this source.}
\label{fig:app1}
\end{figure}

The broad-component emission is seen in the NE galaxy of this galaxy-pair. The broad-component emission is blueshifted and found in the region where the narrow-component disk is redshifted or at the systemic velocity. The broad-component emission appears to originate from the nuclear region, is only seen on one side of the nucleus, and stretches out to $\sim$1.6 kpc. The narrow-component disk shows a 0 \kms\ iso-velocity contour, which position angle (PA) changes from PA\,$\sim$\,0$^{\circ}$ to PA\,$\sim$\,23$^{\circ}$. We argue that the latter is the most likely PA of the minor axis of the galaxy, given that it is perpendicular to the elongated morphology of the galaxy in the HST/NICMOS imaging (Fig.\,\ref{fig:outflows}). This means that the outflow spans a cone that is oriented about 15$-$55 degrees off the minor axis. The outflow peaks in intensity and velocity roughly $\sim$0.5 kpc from the core. Here the outflow has a bulk velocity of $\sim$200 \kms\ with respect to the systemic velocity, which it maintains out to a distance of $\sim$1.6 kpc. The wing of the broad component stretches out to -700 \kms\ at zero-intensity. Table \ref{tab:results} gives more details.
Figure \ref{fig:12112} suggests that the SW galaxy of this ULIRG also shows faint broad-component emission, but Fig. \ref{fig:app1} clarifies that this merely reflects our uncertainty in the Gaussian fitting procedure due to the underlying strong continuum emission of this source.

\subsection{IRAS\,14348-1447} 

The broad-component emission is seen in the SW galaxy of this galaxy-pair, which hosts an AGN that is likely to be Compton-thick \citep{iwa11}. The broad component shows the most blueshifted emission of the four galaxies in our sample, stretching out to -900 \kms\ at zero-intensity. Previous work showed that gas outflows are on average faster in the presence of an AGN \citep[][see also models by \citealt{nar08}]{spo13,arr14,cic14,hil14}. The fact that this is also the case when comparing IRAS\,14348-1447 with the other ULIRGs in our sample strongly suggests that the outflow in IRAS\,14348-1447 originates from the nuclear region. The broad component is resolved only marginally in a westerly direction, with a size of $\sim$0.9 kpc. However, the fact that the narrow-component emission is close to the systemic velocity throughout the disk suggests that the galaxy disk is aligned close to face-on. This makes our estimated PA$\sim$-55$^{\circ}$ for the minor axis, as based on the 0 \kms\ iso-velocity contour and HST/NICMOS morphology, rather uncertain. Nevertheless, if the outflow traced by the broad component were aligned along the minor axis of the galaxy, it would mostly propagate along our line of sight, which means that the true extent and velocity of the broad-component emission may be significantly larger. More details are given in Table \ref{tab:results}.

\subsection{IRAS\,17208-0014} 
\label{sec:app_17208}

IRAS\,17208-0014 has been associated with multi-phase gas outflows by various authors \citep{rup05,mar06,stu11,arr14,sot12,med15}. However, several integral-field studies did not find evidence for such outflows in the ionized phase \citep{arr03,wes12,rup13a}, and \citet{arr03} even argued the presence of gas inflows. In our SINFONI data, the overall H$_{2}$ kinematics appear to follow the CO kinematics of cold molecular gas discussed in detail by \citet{gar15}. The broad-component H$_{2}$ emission is only seen near the systemic velocity, and spreads across $\sim$2\,kpc, which is a substantial fraction of the narrow-component disk. In Sect.\,\ref{sec:17208}, we discuss that this broad-component emission is not likely to be related to an outflow, although its true nature remains unclear. 

We do not detect a hot H$_{2}$ counterpart to the cold molecular outflow previously discovered in CO by \citet{gar15}. As shown in Sect.\,\ref{sec:appA_17208}, we derive a firm upper limit on the outflow of hot molecular H$_{2}$ gas of M$_{\rm H2-hot}$ $\la$400 M$_{\odot}$, which results in a hot-to-cold molecular gas mass ratio of $\la$9\,$\times$\,10$^{-6}$.

\subsection{IRAS\,22491-1808} 

The broad-component emission is seen in the E galaxy of this galaxy pair. The broad component is most prominent and blueshifted west of the nucleus. It spans a cone that is aligned roughly 10$-$65$^{\circ}$ off the minor axis of the narrow-component disk, which we estimate has PA$\sim$50$^{\circ}$, based on the 0 \kms\ iso-velocity contour. A faint redshifted counterpart could be present towards the south, but this needs to be verified. The broad-component emission stretches 0.7\,kpc from the nucleus, where it also shows the highest bulk velocities of roughly -100 \kms\ with respect to the systemic velocity. The maximum velocity of the outflow reaches only -500 \kms\ at zero-intensity. Table \ref{tab:results} and Sect.\,\ref{sec:results} give more details.

\section{Velocity dispersion maps}
\label{sec:app2}

\begin{figure}[h!]
\centering
\includegraphics[width=\hsize]{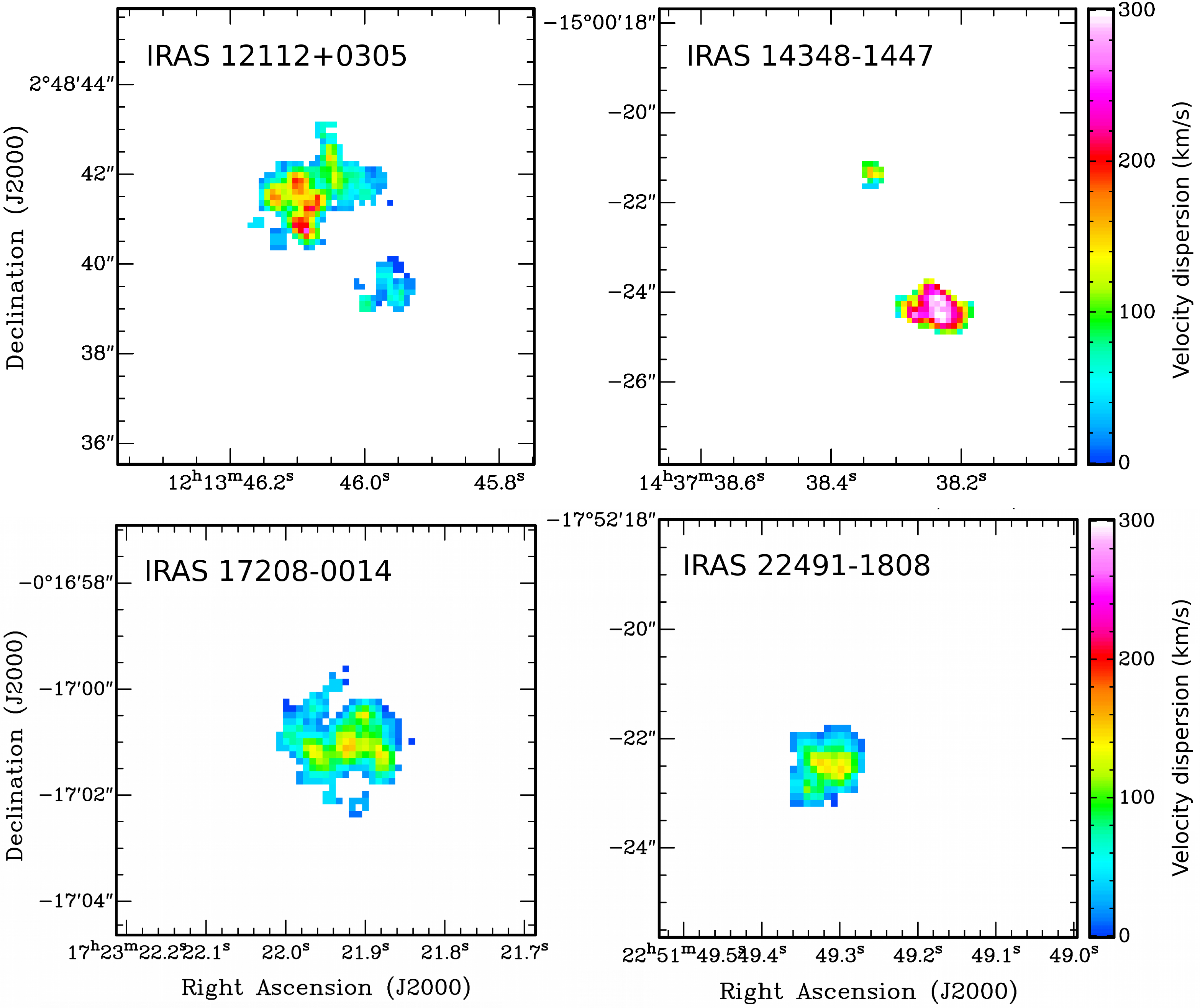}
\caption{Maps of the velocity dispersion, $\sigma$\,=\,FWHM/2.35, of the broad-component H$_{2}$ 1-0\,S(1)  emission in IRAS 12112+0305, IRAS\,14348-1447, IRAS\,17208-0014, and IRAS\,22491-1808. The velocity scaling is the same in all four plots.}
\label{fig:dispersion}
\end{figure}

Figure \ref{fig:dispersion} shows the velocity dispersion of the broad-component emission from Figs.\,\ref{fig:12112}\,$-$\,\ref{fig:22491}. 

\end{appendix}

\end{document}